\documentclass[a4,10pt]{article}

\usepackage{arxiv}

\usepackage[utf8]{inputenc} 
\usepackage[T1]{fontenc}    
\usepackage{url}            
   
\usepackage{nicefrac}       
\usepackage{microtype}      
\usepackage{lipsum}
\usepackage{graphicx}
\graphicspath{ {./images/} }

\usepackage{cite}
\usepackage{amsmath,amssymb,amsfonts}
\usepackage{xcolor}
\definecolor{myorange}{RGB}{100, 50, 0}
\definecolor{myblub}{RGB}{34, 52, 168}
\definecolor{dg}{RGB}{64,64,64}

\usepackage[colorlinks = true,
		citecolor = blue,
		urlcolor= blue,
		linkcolor=dg,]{hyperref}
\usepackage{algorithmic}
\usepackage{graphicx}
\usepackage{textcomp}
\usepackage{xcolor}
\def\BibTeX{{\rm B\kern-.05em{\sc i\kern-.025em b}\kern-.08em
    T\kern-.1667em\lower.7ex\hbox{E}\kern-.125emX}}	

\title{A Systematisation of Knowledge: Connecting European Digital Identities with Web3.}

\date{August 2024}

\author{
Ben Biedermann\\
\href{mailto:bb@acurraent.com}{bb@acurraent.com}\\
Islands and Small States Institute \\
University of Malta\\
Msida, Malta \\
\href{https://orcid.org/0000-0003-1331-6517}{0000-0003-1331-6517}\\
\And
Matthew Scerri\\
\href{mailto:matthew@wid3.org}{matthew@wid3.org}\\
WIDE Consortium \\
Frankfurt/ODER, Germany \\
\href{https://orcid.org/0009-0002-1672-9018}{0009-0002-1672-9018}\\
\And
Victoria Kozlova\\
\href{mailto:vk@acurraent.com}{vk@acurraent.com}\\
ACURRAENT UG \\
Frankfurt/ODER, Germany \\
\href{https://orcid.org/0000-0003-3303-3143}{0000-0003-3303-3143}\\
\And
Joshua Ellul\\
\href{mailto:joshua.ellul@um.edu.mt}{joshua.ellul@um.edu.mt}\\
Centre for DLT \\
University of Malta\\
Msida, Malta \\
\href{https://orcid.org/0000-0002-4796-5665}{0000-0002-4796-5665}\\
  }
 \begin{document}

\maketitle

\begin{abstract}
The terms self-sovereign identity (SSI) and decentralised identity are often used interchangeably, which results in increasing ambiguity when solutions are being investigated and compared. This article aims to provide a clear distinction between the two concepts in relation to the revised Regulation as Regards establishing the European Digital Identity Framework (eIDAS 2.0) by providing a systematisation of knowledge of technological developments that led up to implementation of eIDAS 2.0. Applying an inductive exploratory approach, relevant literature was selected iteratively in waves over a nine months time frame and covers literature between 2005 and 2024. The review found that the decentralised identity sector emerged adjacent to the OpenID Connect (OIDC) paradigm of Open Authentication, whereas SSI denotes the sector's shift towards blockchain-based solutions. In this study, it is shown that the interchangeable use of SSI and decentralised identity coincides with novel protocols over OIDC. While the first part of this paper distinguishes OIDC from decentralised identity, the second part addresses the incompatibility between OIDC under eIDAS 2.0 and Web3. The paper closes by suggesting further research for establishing a digital identity bridge for connecting applications on public-permissionless ledgers with data originating from eIDAS 2.0 and being presented using OIDC.
\end{abstract}


\section{Introduction}
Decentralised identity has served as a catch-all term for various technologies that are prescriptive rather than practical. Self-sovereign identity (SSI) is an example for a value-laden term, used to denote a subcategory of decentralised identity \cite{b1} without explaining what this means for its users. Although decentralised identity and SSI specifically aim to guarantee user privacy and data portability \cite{b2}, the terms and concepts that are used to convey such ends are confusing and ambiguous \cite{b3}. This is further exacerbated by projects spanning across public to private domains.\\

The vast amount of competing and converging projects makes choosing a solution a challenge. The Decentralised Identity Foundation (DIF) alone lists over 180 decentralised identifier (DID) methods \cite{b4}, which reference four archetypes of trust registries, and make use of two major transport layer protocols. Whilst regulators define requirements for regulatory technology, they also report lacking capacity --- not just for digital identity, but also for digital supervisory and regulatory infrastructure at large, such as central bank digital currencies and open banking \cite{b5}. Hence, the risk and cost of integrating a digital identity solution increases, which is further exacerbated by vendor lock-in that comes at the expense of interoperability, regardless of the electronic identification, authentication, and trust services (eIDAS) expert group-specified European Digital Identity Wallet (EUDIW) \cite{b6}, its implementation is still underway. With representatives for each EU member state, the eIDAS expert group united a diverse set of technology stances. Yet, through its resources and supranational position, the eIDAS expert has an advantage over many public and private sector actors within the EU and abroad, when conducting extensive technology evaluations. Moreover, the eIDAS expert group limited permissible trust registries to existing eIDAS-compliant solutions and so called ``electronic ledgers'' \cite{b7}, \cite{b8}. Thus, public-permissionless distributed ledgers continue to be outside the scope of the regulation.\\

As a result, stakeholders of the revised eIDAS regulation, whose operations rely substantially on public-permissionless distributed ledgers do not benefit from the legal certainty provided by Regulation (EU) 2024/1183 \cite{b8}. This paper aims to shed light on address the risk of established interoperability schemes between asymmetric cryptographic key pairs and Open Authentication (OAuth) \cite{b9} becoming obsolete. Therefore, the state-of-the-art review herein asks:\\

\footnotesize
\begin{itemize}
    \item \textbf{RQ1}: How do decentralised identity protocols based on Open Identity Connect (OIDC) differentiate from Web3?

    \item \textbf{RQ2:} How can the EUDIW specification, using novel OIDC protocols, be integrated with existing Web3 applications?
\end{itemize}
\normalsize
In answering the research questions, the paper proceeds as follows. In the next section, the reader is provided with necessary background on the divide between decentralised identity, Web3 adoption, and the eIDAS process in the European Union (EU). Thereafter, the methodology followed is presented. In Section \ref{sec:sota}, the findings are described and structured along two main axes that have emerged over the last twenty years in digital identity research and development. Section \ref{sec:lessons} discusses the results, and the paper closes by drawing conclusions, as well as suggesting further avenues for research.\\

\section{A Brief History of Decentralised Identity}

``Decentralised identity'' was coined in 2007 \cite{b11}. Since then, the protocols that decentralised identity solutions rely on have changed significantly and even the understanding of these protocols shifted multiple times. In 2006, asymmetric cryptography was considered centralised, because for trust to occur it required public key infrastructure (PKI) repositories. The alternative was considered to be OIDC because ``[w]eb-oriented identity protocols such as OpenID [...], which use URIs as basic identifiers [were considered] inherently decentralized systems'' \cite{b12}. The Association for Global System for Mobile Communications (GSMA) defined decentralised identity as ``an emerging concept that gives back control of identity to consumers through the use of an identity wallet'' \cite{b13}, whereas the earliest solutions explicitly labelled as and debated under the term decentralised identity date back to 2005 \cite{b14}.\\

The utilisation of distributed ledger technology (DLT) for guaranteeing the cryptographic integrity of credentials and knowledge of the signer for these credentials lessened the reliance on direct issuance channels between issuers and relying parties \cite{b15}. Subjects of credentials now were able to store, maintain, and present credentials even anonymously without relying on an issuer to do so on their behalf \cite{b16}, \cite{b17}, \cite{b18}. The decentralised identity sector sharply distanced itself from federated identity protocols most widely known as Open Authentication (OAuth) \cite{b19}. This paradigm change occurred with the advent of Ethereum in 2015, which may have caused projects like MicroID to lose traction \cite{b18}, \cite{b20}.\\

Upon the 2017 and 2021 cryptocurrency market downturns, public sector funding for decentralised identity solutions became more prominent. This is shown by DLT-based identity solutions shifting to participate in EU funded projects, such as \textit{ESSIF Labs} \cite{b21}.Not only the European Blockchain Service Infrastructure (EBSI)\footnote{Now part of EUROPEUM-EDIC. \href{https://digital-strategy.ec.europa.eu/en/news/blockchain-creation-europeum-edic}{digital-strategy.ec.europa.eu/en/news/blockchain-creation-europeum-edic}} took shape, but also the Next Generation Internet (NGI) programme\footnote{\url{https://www.ngi.eu/}} as well as the Government of British Columbia invested heavily into decentralised identifiers (DIDs) from 2018 on \cite{b21}, \cite{b22}, \cite{b23}. Meanwhile, the EU funding for consolidating the divergent technology development paths of OIDC and DLT does not match the outcome of the eIDAS-process, which left blockchain-based solutions out.\\

\section{Methodology}

This paper provides a brief state-of-the-art (SOTA) review to ascertain the reasons for the divergence between OIDC-based identity solutions and decentralised identities using public-permissionless blockchains. In spite of SOTA reviews being methodologically under-defined \cite{b24}, the evolution of the digital identity divide follows a six step SOTA method of Barry et al. \cite{b25}. In the first step, the digital identity divide was identified as the field of interest and inductive selection-based document analysis was established as the appropriate research method \cite{b26}. As a second step, the timeframe of the analysis was determined to span from approximately 2005 to 2024. The third step finalised the research questions mentioned in the introduction. Thereafter,  step four consisted of defining the search strategy for responding to the research question. Given the velocity of changes to draft standards for novel OIDC protocols at the Internet Engineering Task Force (IETF) and ongoing changes to the EUDIW architecture reference framework (ARF), exploratory forward and backward search of cited literature combined with continuous searches on Google Scholar, was identified as the most practical research structure. Fifth, the selected literature was reviewed and relevant data was extracted. Finally, the process was iterated to arrive at succinct and consistent results.\\

More concretely, Google Scholar searches were performed over a period of four months starting in November 2023 and ending in the beginning of March 2024. The searches were motivated by the aim for defining decentralised identity, which has proven difficult as most research that emerged in the first two result pages for Google Scholar for the search string \textit{"decentralised identity" OR "decentralized identity"} did not define the term. Instead, the term \textit{decentralised identity} was used to describe a specific digital identity implementation. This prompted the research to shift from a mere definition of \textit{decentralised identity} to undertaking a SOTA analysis that deepened the preliminary results of the inductive research that showed both OIDC-based and DLT-enabled solutions using the term \textit{decentralised identity}.\\

Following, paper pertaining to the first research question were grouped as "Web3" or "OIDC" and resulted in the selection of five scholarly contributions out of twenty search results, based on their title, abstract, and year of publication. As a SOTA review is "time-based and turning point-based" \cite{b24}, historic clustering was introduced as a second aspect of the methodology. To that end, the research context of eIDAS 2.0 and the ARF, as well as their limitations in accounting for existing DLT-based solutions. Therefore, the research proceeded abductively, posing RQ2. From the initial set of selected papers, three \textit{turning-points} were identified, \textit{i.e.} 2005 -- 2009, 2014 -- 2018, and 2020 -- 2024. The search then proceeded to select articles that were cited or citing these works and were published during these timeframes. Afterwards, additional Google Scholar searches for \textit{("OIDC" OR "OpenID Connect") AND ("EUDIW" OR "European Digital Identity")} were performed and results were filtered according to the relevance criteria and the time-frames. Lastly, both authors' affiliations and the examples in the total twenty selected scholarly contributions were matched against an existing self-compiled database. This database comprises 178 private sector digital identity projects and 413 national digital identity schemes.\\

\section{State-of-the-Art of the Digital Identity Divide}
\label{sec:sota}

The cases of Malta and Germany highlight what is at stake, when excluding blockchain-based digital identity solutions from (EU) 2024/1183 and the ARF. Given Malta’s specialisation on the regulation of virtual financial assets \cite{b27}, i.e. digital assets on public blockchains, the inclusion of decentralised identity over public-permissionless distributed ledger in eIDAS 2.0 could have closed the gap between local virtual financial asset regulation, the Directive on Markets in Crypto-Assets, and the first eIDAS regulation. Accordingly, the Malta Digital Innovation Authority (MDIA) strategy for 2023 to 2025 explicitly mentions the implementation of an European Digital Identity Wallet (EUDIW) as a core policy objective \cite{b28}. Conversely, as a federal state, Germany has diverse requirements for an EUDIW, where a greater diversity of solutions can reduce technology risk, but increase interoperability issues. For this reason, a public consultation process was launched in Germany that accounts for required technological plurality in responding to the EUDIW specification.\\

\begin{table}[htbp]
\begin{center}
\caption{Relevant identity projects across the (de)centralised divide}
\renewcommand{\arraystretch}{1.5} 
  \footnotesize
    \begin{tabular}{p{2.2cm}p{5.6cm}}
    \bfseries{Project Name} & \textbf{Technology} \\
    \hline
    ClaimID & OAuth 2.0, SHA-1 hashed email addresses \\
    Iden3 & Ethereum, JSON-LD \\
 Disco.xyz & Ethereum, JSON Verifiable Credential Data Model, DID:3/ WEB/ PKH \\
    BC Digital Trust & Hyperledger Indy, Hyperledger Aries Cloud Agent Python (ACA-Py), AnonCreds \\
    NGI ESSIF Lab & cross-platform EBSI, Hyperledger Indy, KERI, x.509 \\
    Smart eID & TR-30110, Samsung Secure Element \\
    German EUdiW & TR-30110, OID4VCI, OID4VP, SIOP, SD-JWT, mdoc \\
    EUDIW ARF & OID4VCI, OID4VP, SIOP, SD-JWT, mdoc \\
    EBSI & Verifiable Credential Data Model, DID:EBSI \\
    Malta eID & x.509 Certificates, REST API \\
 \end{tabular}
\label{tab1}
\end{center}
\end{table}

\subsection{Review of Decentralised Identity Frameworks}

While the most widely used credential format, anonymous credentials, drew heavily on the DLT infrastructure of Hyperledger \cite{b30}. The need for trustworthy, un-spoofed, and valid hashes of public keys prompted the latest solutions under the decentralised identity paradigm to resort to using public-permissionless distributed ledgers \cite{b18}, \cite{b19}, \cite{b31}, issuer signed decentralised identifier (DID) documents on content addressed storage solutions like the Interplanetary Filesystem (IPFS) \cite{b32}, or centralised verifiable data registries (VDRs), such as public key infrastructures (PKIs) \cite{b33}.\\

Ultimate control over the key pair that is used to establish secure connections with counterparties and in most cases binds the credential data to a subject, determines the degree of decentralisation of identity access management (IAM). There is a trend to re-centralise the key management of decentralised identity schemes by hosting key pairs hosted on a server, turning decentralised identity solutions into federated systems \cite{b34}. This is evident in the emergence of SSI Cloud Wallets \cite{b35}, \cite{b36}, \cite{b37}. In these cases, SSI-based software as a service (SaaS) infrastructures only replicate Web 2.0 architectures. Contrary, server-side credential management in decentralised identity does not necessarily re-centralise applications. Notable decentralised applications (dApps) that fall in this category are Disco.xyz or iden3 \cite{b38}, \cite{b39}, both relying on blockchains. Due to their usage of DLTs and content addressed storage, they also pose specific challenges for GDPR-compliance \cite{b32}. In this regard, any cloud wallet, even under the paradigm of decentralised identity, prompts additional privacy considerations.\\

\subsection{The State of Digital Identity as Regulatory Technology}

For the purposes of reconciling blockchain-based decentralised identity with the technologies used in the EUDIW, the SOTA review included the European legislative process. After substantial revisions \cite{b40}, the European Commission's proposal for amending the first eIDAS regulation \cite{b41}, \cite{b42} was adopted as Regulation (EU) 2024/1183. The procedure is also known as eIDAS 2.0 \cite{b43} and was included in the SOTA review for its significance for digital identity standards, as well as the relevance to solve lacking Web3 compatibility in the reference states.\\

Generally, the EUDIW allows users to store, access, and manage their governmentally issued credentials, such as national identities, health data, or educational credentials \cite{b44}, \cite{b45}, \cite{b46}, \cite{b47}, \cite{b48}. For enabling this multitude of use cases, the ARF draws from the existing eIDAS regulation and the trust framework that is built upon the legal basis of 910/2014/EU \cite{b49}. Thus, issuances of verifiable credentials \cite{b50} to the EUDIW and the lower-level data that are attested to by a third party, will rely substantially on the centralised PKI-based asymmetric cryptography with its respective identifiers that was established through or alongside 910/2014/EU. In COM/2021/281 these attestations are referred to as (qualified) electronic attribute attestations ([Q]EAAs), which indicates a division between attributes issued by so-called notified trust service providers \cite{b42} and EAA issued by private or public third parties.\\

The decentralised identity suite by the OpenID Foundation utilises established token-based authentication with JSON web tokens (JWTs) by combining them with selective disclosure functionalities \cite{b51}, \cite{b52}. Selective disclosure is important for enabling predicates, a sought after feature due to its promises of data minimisation \cite{b53}. Meanwhile, JWT for selective disclosure (SD-JWTs) does not use experimental cryptography that is capable of zero-knowledge proofs (ZKPs) \cite{b51}, such as BBS+ \cite{b54}. To overcome the lack of ZKPs, SD-JWTs split the VCs in multiple claims, and issue several VCs for individual claims in plain text and as salted hashes to achieve attribute blinding and mitigate correlation by verifiers.\\

As a result, the OIDC standards for decentralised identity are available with JSON web signatures (JWS) for signing the credential, which are compatible with the named curves published by the National Institute of Standards and Technology (NIST), such as the NIST P-Series and Brainpool series \cite{b55} recommended by the European Technical Standards Institute (ETSI) for digital signature schemes \cite{b56}. EU member states have turned these advantages, however, into complacency with national electronic identity schemes that are in production.\\

Even for the purposes of the EUDIW, SD-JWT does not provide sufficient functionality. Thus, the ARF requires EUDIWs to support the standard for the mobile driving licence (mDL) \cite{b7}. The mDL relies on so-called MDOCs for binding credential data to a public key \cite{b57}. In turn, the mdoc draws from the jointly developed technical guideline TR-30110 for key provision in electronic identity schemes by the German Federal Office for Information Security (BSI) and the French Cybersecurity Agency (ANSSI) \cite{b58}. Thus, the ARF includes exclusively innovations of established and centralised digital identity schemes that are incompatible with the cryptography used by blockchains and advanced decentralised identity schemes.\\

\section{Lessons for EU Digital Identity Bridges}
\label{sec:lessons}
The SOTA review highlights that decentralised identity solutions have come full circle. The digital identity sector today returned to the origins of decentralised identity solutions based on OIDC transport protocols and centralised trust architectures for asymmetric cryptographic keys. The use of DLT, with blockchains in particular, by and large has been marginalised on the European level. Credential trust and holder binding, as well as protocols for credential transmission that do not rely on centralised trust require an additional interoperability layer for constructing, presenting, and managing credentials in a Web3-compatible way.\\

In this regard, a bridging solution that aggregates, encrypts, and associates credentials of all sorts, including from OIDC protocols, may be necessary to enable Web3 use cases of personal identification data (PID). The decentralised identity solution SOTA suggests that it is beneficial to store user key pairs in a Web3 wallet solution on their device and the user key encrypted credentials – on a cloud-based storage solution that does not have to be trusted. Furthermore, this approach follows best practices for the issuance of predicates described in the draft specification for SD-JWTs and offers a privacy-preserving alternative to OIDC-based SD-JWTs that are specified in the ARF. In essence, the digital identity bridge is based on predicates or salted hashes of claims by default, mimicking in SD-JWTs. Instead of presenting a PID in plain text whenever a credential is presented to a relying party, the bridging solution only uses the Web3 public key of the user. Thus, the required bridging solution reduces privacy risks of EUDIW according to the ARF. In other words, the SOTA with respect to connecting European digital identities with Web3 can be furthered by establishing interoperability, mitigating user privacy concerns, and alleviating user friction caused by fully decentralised identity applications.\\

\section{Limitations}

The approach proposed based on the results from the SOTA review, however, has several limitations. First, its focus on enhancing the availability of EUDIW data to Web3 trades of compliance with eIDAS 2.0. Thus, the bridge cannot be used for governmental use cases requiring a the presentation of a PID through level-of-assurance High. Second, it may be seen as yet another digital identity scheme by relying parties and issuers. Third, in its current stage it does not offer crypto agility. Nevertheless, these limitations are of a technical nature and can be addressed through further development. The importance of the proposal lays in the concept and its technological neutrality, emphasising a privacy-first approach.\\

\section{Conclusions}

The use of users’ public keys to encrypt the bridged data and retrieved on demand, conceptually outright contrasts architectures prevalent in both the decentralised and centralised digital identity sector. Based on the SOTA review, it is an innovative approach that appears to have the potential of reconciling the divergent technology stacks for the European digital identity framework and decentralised identity. Moreover, a hybrid approach for storing user key pairs and credentials allows for connecting Web3 with EUDIWs. Given that this proposed approach is grounded in the exploratory SOTA review, it indicates that digital identity solutions in Web3 differ from the EUDIW in two main respects. First, Web3 solutions enable the use of advanced cryptography that extends the SOTA, which is present in the technical standards on national and European level. Second, Web3 identity ``[p]rotocols [allow] for trustworthiness of entities by means of verifiable credentials and decentralized reputation systems'' \cite{b58}.\\

Despite the fact that the ARF and the Regulation (EU) 2024/1193 explicitly exclude blockchains from becoming trusted ledgers \cite{b7}, \cite{b8}, abandoning the notion of SSI in favour of identity abstraction services could integrate credentials from OIDC-based EUDIWs. Rather than conducting research for the development of new trust infrastructures, which potentially further fragment both the European digital identity landscape and decentralised identity, the SOTA suggests that Web3 interoperability requires solutions that recombine existing standards and specifications. The SOTA enables novel compliance and engagement flows through the use of zero-knowledge attestation. Additionally, hybrid credential storage of encrypted user data, which is more pragmatic than existing SSI architectures. Hence, the decentralised identity sector has moved beyond dogmatic principles of early SSI implementations, such as Cameron (2007) and  Allen (2016) have put forth \cite{b2}, \cite{b10}.\\

In turn, decentralised identity today offers flexibility in either choosing established elliptic curve cryptography, such as the signature algorithms developed for the Edward’s curve 25519 \cite{b59} or allowing selective disclosure through ZKPs \cite{b54}. Therefore, identity bridging enables sector-specific adoption of novel identification mechanisms into on-ramping and off-ramping flows and could address uncertainty of using virtual asset service providers (VASPs) \cite{b60}. For this reason, further research is needed to enhance the understanding of the processes for the development of bridging solutions that emphasise trustworthiness, reliability, and integrity.\\

\section*{Acknowledgment}

We are grateful to our reviewers for their insightful remarks and our research communities for their feedback on early drafts of this work. This work has been funded by the EU in the framework of the NGI TRUSTCHAIN project, grant number: \href{https://cordis.europa.eu/project/id/101093274}{101093274}.

\end{document}